\documentclass[conference]{IEEEtran}

\ifCLASSINFOpdf
\else
\fi

\usepackage{graphicx}
\usepackage[cmex10]{amsmath}
\usepackage{fixltx2e}
\usepackage{booktabs}
\usepackage{glossaries}
\usepackage[ruled]{algorithm2e} 
\usepackage{algpseudocode}
\usepackage{xcolor}
\usepackage[export]{adjustbox}
\usepackage{mathtools}
\usepackage{cite}
\usepackage{placeins}
\usepackage{stfloats}
\usepackage{amsmath}

\SetKwRepeat{Do}{do}{while}

\SetKwRepeat{Do}{do}{while}%

\begin{document}

\title{Bio-Inspired Resource Allocation for Relay-Aided Device-to-Device Communications}

\author{\IEEEauthorblockN{Christoforos Vlachos\IEEEauthorrefmark{3}, Hisham Elshaer\IEEEauthorrefmark{1}\IEEEauthorrefmark{3}, Jian Chen\IEEEauthorrefmark{2},
Vasilis Friderikos\IEEEauthorrefmark{3} and Mischa Dohler\IEEEauthorrefmark{3}} 
\IEEEauthorblockA{\IEEEauthorrefmark{3}Centre for Telecommunications Research, King's College London, UK.
\IEEEauthorblockA{\IEEEauthorrefmark{1}Vodafone Group R\&D, Newbury, UK. } 
\IEEEauthorblockA{\IEEEauthorrefmark{2}Northeastern University, China. } 
\textit{Email}: \{christoforos.vlachos, vasilis.friderikos, mischa.dohler\}@kcl.ac.uk\\
hisham.elshaer@vodafone.com, chenjian@mail.neu.edu.cn}

}

\maketitle

\begin{abstract} 
The Device-to-Device (D2D) communication principle is a key enabler of direct localized communication between mobile nodes and is expected to propel a plethora of novel multimedia services. However, even though it offers a wide set of capabilities mainly due to the proximity and resource reuse gains, interference must be carefully controlled to maximize the achievable rate for coexisting cellular and D2D users.  
The scope of this work is to provide an interference-aware real-time resource allocation (RA) framework for relay-aided D2D communications that underlay cellular networks. The main objective is to maximize the overall network throughput by guaranteeing a minimum rate threshold for cellular and D2D links. To this direction, genetic algorithms (GAs) are proven to be powerful and versatile methodologies that account for not only enhanced performance but also reduced computational complexity in emerging wireless networks. Numerical investigations highlight the performance gains compared to baseline RA methods and especially in highly dense scenarios which will be the case in future 5G networks.        
\end{abstract} 

\begin{IEEEkeywords}
Device-to-Device communications, underlay, relay, resource allocation, genetic algorithm, 5G.
\end{IEEEkeywords}

\IEEEpeerreviewmaketitle

\section{Introduction}

Device-to-device (D2D) communication emerges as an attractive way to tackle the dramatic increase in traffic and shortage of spectrum in cellular networks by capitalising on the proximity of user equipments (UEs) to each other. This enables the direct communication between two or more UEs without the need for routing the data flows conventionally through the cellular base station (BS) \cite{Lin2014}. D2D opens the door to a multitude of proximity based applications such as public safety,  peer-to-peer communication, local advertisement, multi-player gaming and a lot more. 

D2D as an underlay in cellular networks enables the reuse of the spectrum assigned for cellular communications. It also allows the offloading of cellular traffic and enables more reliable and high throughput links between users in close proximity. For this reason, and following the prediction for further densification in future 5G networks, D2D is expected to play a principal role in spectrum and resource management since in several cases the number of D2D connections can be very high and the resources would need to be carefully controlled. However, in a D2D enabled network some challenges need to be addressed in order to get the full benefit of this technology. Firstly, the potential D2D UEs may not be in close proximity which may render the establishment of a reliable connection between the D2D UEs challenging. In addition, the high spectral efficiency of underlay operation comes at the price of high levels of interference to and from cellular UEs (CUEs) which could jeopardize the quality of service (QoS) of D2D as well as cellular UEs.

\subsection{Related work}
Bio-inspired genetic algorithms (GAs) \cite{Johnson95} have become a popular approach in solving resource allocation problems in wireless networks \cite{ZheGua10, LimAra07, TaoCha06} mainly because of their versatility, scalability and computational simplicity which make GA a very attractive method to solve the resource allocation problem as will be detailed in Section \ref{genetic}.
Resource allocation for D2D communications has been extensively studied within the literature. In \cite{LemDow15}, a proportionally fair utility maximization approach is used to allocate resources to both D2D UEs (DUEs) and cellular UEs (CUEs). In  \cite{LinYus13} the mode selection and resource allocation problems for underlay D2D communication are investigated and solved using particle swarm optimization. Further, an efficient graph-theoretical approach is proposed in \cite{MagSta15} to perform channel allocation for DUEs. Resource allocation in relay-aided D2D scenario has been studied in \cite{HasHos14}.

\subsection{Contribution}
In this paper, we study the joint resource allocation for cellular and relay-aided underlay D2D communications where DUEs share the UL resources with CUEs. We consider that a UE could act as relay node in order to enhance the link quality between DUEs that are far apart or the channel quality between them is poor. All DUEs are eligible to either communicate directly with their peer or via a relay. Normally, relays are used to improve network coverage where it is needed. However, in order to offload the traffic that should traditionally be routed via the BS, a relay can also become the intermediate node that assists two UEs to communicate, without adding extra burden on the BS side. In such a scenario, our proposal considers the use of GAs in order to find a near-optimal allocation of resources for CUEs and DUEs that can achieve the maximum sum-rate. We compare the GA performance with a heuristic algorithm that prioritizes the D2D resource allocation as well as with a random allocation scheme. Differently from \cite{HasHos14} which considered that all traffic flows are routed through L3 standard relays, in our study the choice of direct or relayed D2D communication based on the achievable rate is part of the optimization problem.

To the best of our knowledge, this is the first work that combines the optimization of the mode selection between direct and relayed D2D operation with the aim to achieve a joint resource allocation of cellular and D2D communication that maximizes the aggregate throughput.

\section{Problem Definition}
\label{prob_def}
The resource allocation problem in cellular networks is a widely studied area that falls within the nature of NP-hard problems which cannot be solved in real-time. Popular integer relaxation methods have been applied to reduce its time complexity but do not render it a real-time solution for network operators. In this section we define important preliminary notations and parameters that will help us further formulate the relay-aided D2D/cellular resource allocation optimization setting and pave the way for our proposal.  

\begin{figure}
\centering
\includegraphics[width=0.47\textwidth, trim = .5cm .2cm .4cm .5cm,clip = true]{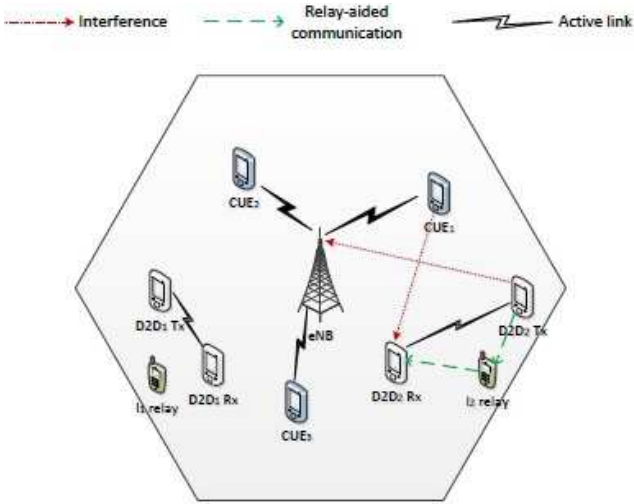}
\caption{Uplink scenario of relay-aided D2D communications as an underlay to the cellular network.}
\label{fig:scenario}
\end{figure}

\subsection{Preliminaries}
First, we consider the uplink (UL) case scenario of D2D underlaying a cellular network where interference patterns from a CUE to a receiving DUE and from the transmitting DUEs to the CUE UL transmission, as depicted in Fig. \ref{fig:scenario}. In this figure, interference exerted from the cellular user $CUE_1$ towards the $D2D_2$ pair and vice versa might be destructive not only for the reliability of the link, but also for the aggregate network throughput. Therefore, these two transmissions should occupy different Resource Blocks (RBs) to avoid mutual interference. Additionally, we assume that cellular users are directly transmitting to the serving BS, whereas the communication mode between two DUEs can be either direct or with the help of a closely located relay that lays within the geographical serving area of a macro cell. An important assumption is that only one proximate UE to a D2D pair can be used as its relay to help the transmission.  
Now, before we detail the problem formulation, we need to define the following sets:  

\begin{itemize}
\item $\mathcal{N} = \{1, 2, \hdots, N\}$: set of available RBs. 
\item $\mathcal{D} = \{1, 2, \hdots, D\}$: set of D2D links. 
\item $\mathcal{C} = \{1, 2, \hdots, C\}$: set of cellular links. 
\item $\mathcal{L} = \{1, 2, \hdots, L\}$: set of relays. 
\end{itemize}

Also, in order to formulate this problem, we need to further define the decision variables of the optimization setting that will be valued according to the assignment (or not) of a RB to a specific user, either for a cellular, a direct or relayed D2D communication. The binary variables that correspond to CUEs, relayed D2D or direct D2D RB allocation are defined by (\ref{eq:dec_var1})-(\ref{eq:dec_var3}) respectively.

\begin{equation}
\label{eq:dec_var1}
x_{c}^n = 
\begin{cases}
1, & \text{if CUE } c \in \mathcal{C} \text{ transmits on RB } n \in \mathcal{N} \ \ \ \ \ \\  
0, & \text{otherwise.}
\end{cases}
\end{equation}
\begin{equation}
\label{eq:dec_var2}
y_{ij}^n = 
\begin{cases}
1, & \text{if DUE (relay) } i \text{ sends to relay (DUE) } j \text{ via } n  \\  
0, & \text{otherwise.}
\end{cases}
\end{equation}
\begin{equation}
\label{eq:dec_var3}
z_{d}^n = 
\begin{cases}
1, & \text{if D2D pair } d \text{ communicates directly with RB } n  \\  
0, & \text{otherwise.}
\end{cases}
\end{equation}

We consider a deterministic model where the signal-to-interference-and-noise-ratio (SINR) between nodes $i$ and $j$ over RB $n$, denoted as $\gamma_{ij}^n$, can be expressed as
\begin{equation}
\label{eq:rec_power}
\gamma_{ij}^n = \frac{P_i G_{ij}}{I_{j,n} + N_0}, 
\end{equation}
where $I_{j,n}$ is the interference received by user $j$ over resource block $n$, $P_i$ is the transmitted power of node $i$, $G_{ij}$ is the link gain between node $i$ and $j$, and lastly, $N_0$ is the power of background/thermal noise. 
The D2D interference to the UL transmission of CUE $c$ to BS $b$ over RB $n$ is denoted by $I_{c_b,n}$ and is given by
\begin{equation}
\label{eq:bs_interf}
I_{c_b,n} = \sum_{d \in \mathcal{D}} \bigg( P_d G_{db} z_d^n  + \sum_{l \in \mathcal{L}} (P_{d} G_{db} +  P_l G_{lb}) y_{dl}^n \bigg).
\end{equation}
We note that, for the rest of the paper, the $i, j$ indexes in $G_{ij}$ (or $y_{ij}$) correspond to the transmitter and the receiver respectively. Also, from now on, we use $y_{dl}$ to refer to the link between a DUE of pair $d$ to relay $l$ and vice versa. 

The uplink channel rate of the cellular user $c$ over resource block $n$, denoted by $R_c^n$, is given by
\begin{equation}
\label{eq:cell_rate}
R_c^n = B\log_2\bigg(1 + \frac{P_c G_{cb}x_c^n}{I_{c_b,n} + N_0}\bigg),
\end{equation}
where $B$ is the RB bandwidth (180 kHz), $P_c$ is the transmit power of CUE $c$.
Finally, the overall data rate for this CUE is
\begin{equation}
\label{eq:cell_rate_tot}
R_c = \sum_{n \in \mathcal{N}} R_c^n.
\end{equation}
Similarly, the interference affecting the D2D receiver of a pair $d$ can be from the cellular user $c$ or the other DUEs/relays that are transmitting over the same resource. If the resource block $n$ is assigned to $d$, the received interference power for $d$, denoted by $I_{d,n}$, is given by
\begin{equation}
\begin{split}
\label{eq:d2d_interf}
I_{d,n} = \sum_{c \in \mathcal{C}}  P_c G_{cd} x_c^n ~+ \sum_{i \in \mathcal{D} \setminus \{d\}} \Bigg(  P_{i} G_{id} z_{i}^n \\ + \sum_{l_i \in \mathcal{L}} (P_i G_{id} +  P_{l_i} G_{l_id}) y_{il_i}^n \Bigg).
\end{split}
\end{equation}
The rate of the direct D2D communication of link $d$ is then given by
\begin{equation}
\label{eq:d2d_rate}
R_{\text{direct},d} = B  \sum_{n \in \mathcal{N}}\log_2\bigg(1 + \frac{P_d G_{d} z_d^n}{I_{d,n} + N_0}\bigg),
\end{equation}
where $G_d$ is the channel gain for the D2D pair $d$ transmission.

If the relay-based communication is used for D2D pair $d$ via a relay $l$, the link capacity of the first and second hop respectively are given by
\begin{equation}
\label{eq:relay_rate_1}
R_{dl}^{n} = B\sum_{n \in \mathcal{N}}\log_2\bigg(1 + \frac{ P_d G_{dl} y_{dl}^n}{I_{l,n} + N_0}\bigg),
\end{equation}
\begin{equation}
\label{eq:relay_rate_2}
R_{ld}^{n} = B\sum_{n \in \mathcal{N}}\log_2\bigg(1 + \frac{P_{l} G_{ld} y_{ld}^n}{I_{d,n} + N_0}\bigg),
\end{equation}
where $I_{l,n}$ is defined as the interference power from CUE and the other D2D users exerted to relay node $l$, and is given by exchanging the subscript $d$ by $l$ in (\ref{eq:d2d_interf}). 

Lastly, if we consider that relays are operating in full-duplex (FD) mode in amplify-and-forward communication \cite{Riihonen2009}, $R_{l}^{n}$, which is given below, denotes the total achieved rate for a relay-aided D2D communication over RB $n$ where $l$ refers to the relay that assists the considered D2D pair $d$.
\begin{equation}
\label{eq:total_rel_rate}
R_{l}^{n} = \text{min}\{ R_{dl}^{n}, ~R_{ld}^{n} \}.
\end{equation}
\subsection{Problem formulation}
\label{prob_form}
We define the sum-rate maximization problem in an LTE scenario where D2D UEs underlay cellular communications:
\begin{equation}
\label{eq:centralized_optim}
\text{max} \:\: \sum_{n \in \mathcal{N}}  \Bigg[ \sum_{c \in\mathcal{C}} R_{c}^n x_{c}^n + \sum_{d \in\mathcal{D}} \bigg( R_{\text{direct},d}^{n} z_{d}^n + \sum_{l \in \mathcal{L}} R_{l}^{n} y_{dl}^n \bigg) \Bigg]
\end{equation}
\begin{IEEEeqnarray}{ll}
\text{s.t.} ~ 
\sum_{n \in \mathcal{N}} R_{c}^n x_{c}^n \geq R_{th},  ~ \forall c \in \mathcal{C} \IEEEyessubnumber \label{eq:con1}\\
\sum_{n \in \mathcal{N}} \bigg( R_{\text{direct},d}^{n} z_{d}^n + \sum_{l \in \mathcal{L}} R_{l}^{n} y_{dl}^n \bigg) \geq R_{th}, ~ \forall d \in \mathcal{D} \IEEEyessubnumber \label{eq:con2}\\
\sum_{c \in \mathcal{C}} x_{c}^n = 1, ~ \forall n \in \mathcal{N} \IEEEyessubnumber \label{eq:con4in}\\
\sum_{n \in \mathcal{N}} x_{c}^n = 1, ~ \forall c \in \mathcal{C} \IEEEyessubnumber \label{eq:con4}\\
\sum_{n \in \mathcal{N}} \bigg( z_{d}^n + \sum_{l \in \mathcal{L}} y_{dl}^n \bigg) = 1,  ~ \forall d \in \mathcal{D} \IEEEyessubnumber \label{eq:con7}\\
x_{c}^n, y_{dl}^n, z_{d}^n \in \{0,1\}, ~ \forall n \in \mathcal{N}, ~d \in \mathcal{D}, \nonumber \\ l \in \mathcal{L}, ~c \in \mathcal{C}. \IEEEyessubnumber \label{eq:con_last}
\end{IEEEeqnarray}

Constraints \eqref{eq:con1}, \eqref{eq:con2} restrict the rate to be above a predefined threshold for all communications, i.e. direct, relayed D2D and cellular connections. Following the milestones of LTE, \eqref{eq:con4in} imposes the orthogonal assignment of the cellular users.  Also, constraint \eqref{eq:con4} signifies the allocation of each cellular user $c$ with a single RB, whereas \eqref{eq:con7} applies the same RB limitation for the D2D communication and also implies that only one relay can be potentially assisting each D2D link. Thus, the role of the binding variables $z, y$ in the latter constraint is to restrict each D2D to communicate only in direct or relay mode and can be considered as a logical OR set of constraints. 

\section{Genetic Algorithm}
\label{genetic}

GA is one of the most popular bio-inspired algorithms and is used to tackle real world NP-hard optimization problems. In general, bio-inspired algorithms imitate the natural evolution of biological organisms to provide a robust, near optimal solution for various problems. GA is inherently an evolutionary process that involves chromosome encoding, population initialization, fitness function depiction, crossover and selection mechanisms. These operations will be briefly explained in Section \ref{GA_operation}. A detailed analysis of GAs can be found in \cite{Johnson95}. Initially, we introduce the following two important definitions.

\textbf{Problem mapping}:
The first step in solving the resource allocation problem using GA is to establish a mapping between them.
Since our problem space corresponds to CUE or DUE channel allocation, an integer based chromosome coding mechanism will be used. 
Based on this, each individual can directly map to a potential channel allocation for CUEs and DUEs where a channel allocation for a UE is represented by a chromosome; a set of chromosomes forms an individual. The initial population consists of a certain number of individuals, denoted by $M$. A common method to initialize the population is to randomly generate the chromosomes of each individual. In addition, the feasibility of each individual should be ensured to accelerate the convergence process. Thus, we first randomly generate two feasible vectors for each node, according to the representation scheme. Once all vectors are available, they will be combined to form a feasible individual with length equal to $(C+D+L)$. This is repeated until $M$ individuals are generated. The formed population then acts as the very first generation that kicks off the subsequent evolving steps.

\begin{equation} \label{eq:fit_func}
\begin{split}
f = \sum_{n \in \mathcal{N}}  \Bigg[ \sum_{c \in\mathcal{C}} R_{c}^n x_{c}^n + \sum_{d \in\mathcal{D}} \Bigg( R_{\text{direct},d}^{n} z_{d}^n + \sum_{l \in\mathcal{L}} R_{l}^{n} y_{dl}^n \Bigg) \Bigg] \\ + \sum_{c \in\mathcal{C}}\alpha_1 \text{min} \Bigg( R_{th} - \sum_{n \in \mathcal{N}} R_{c}^n x_{c}^n, 0 \Bigg) \\ + \sum_{d \in\mathcal{D}}\alpha_2 \text{min}\Bigg(R_{th} - \sum_{n \in \mathcal{N}} \bigg( R_{\text{direct},d}^{n} z_{d}^n + \sum_{l \in\mathcal{L}} R_{l}^{n} y_{dl}^n \bigg), 0 \Bigg)
\end{split}
\end{equation}
 
\textbf{Fitness function}:
To this end, we firstly need to interpret the objective of the optimization problem in \eqref{eq:centralized_optim} to a fitness function that evaluates the quality of a given individual. In this case, to formulate this we apply a penalty function to ensure that constraints \eqref{eq:con1} - \eqref{eq:con2} are satisfied. In addition, the D2D mode selection (i.e. direct or relayed) is also optimized during fitness evaluation. The fitness function is defined in (\ref{eq:fit_func}).

\subsection{GA operation}
\label{GA_operation}
\textbf{1) Selection}: An operation used for choosing individuals to participate in reproduction. In this study, the roulette wheel selection model is used where the chosen probability is proportional to the individual fitness evaluation function. Its selection probability for individual $i$ is defined as 
\begin{equation}
p_i = \frac{f(i)}{\sum_{i \in M} f(i)}.
\end{equation}

\begin{figure}
\centering
\includegraphics[scale=0.42]{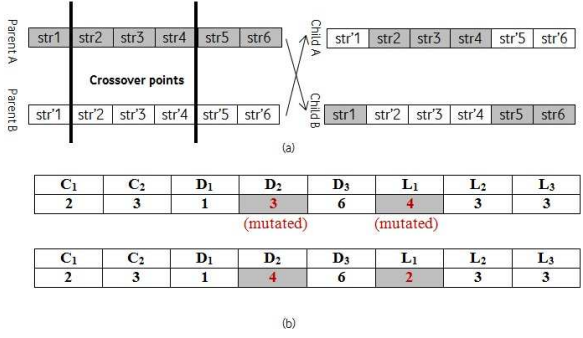}
\caption{(a) Two-point crossover example. (b) Mutation example.}
\label{fig:Cross_mut}
\end{figure}

\textbf{2) Crossover and mutation}: Crossover mixes the current solution so as to find better ones whereas mutation helps the GA  avoid local optima. We use one and two points (OP and TP) crossover cases in our results for comparison. An example of a two point crossover is illustrated in Fig. \ref{fig:Cross_mut}(a). The mutation operation works by randomly making minor changes in the chromosomes after the crossover operation is performed. In our algorithm, we view each chromosome as a single gene.  We define a trivial probability $p_m$ as the likelihood of a gene to mutate. If a gene is determined to mutate, one digit of the vector will be randomly selected and replaced with a different value as shown in Fig. \ref{fig:Cross_mut}(b). 

\textbf{3) Replacement}: After generating a new population, an elitist based replacement model is employed to modify the old population with a certain number of new individuals. The worst individuals in the parental population are replaced by their children in the next generation.

The algorithm works as follows: an initial population is initialized. Then, the reproduction process starts, including mutation and crossover. The worst individuals are replaced with fitter ones based on the fitness function and this process is repeated until the maximum number of generations is reached. Considering the run-time performance of the GA, it is dependant on the three mentioned procedures. It is proven that GA scales well in terms of time complexity compared to ILP that are unable to run for highly dense topologies \cite{Badia2009}.

\begin{algorithm}
\label{alg:heuristic_alg}
\SetKwInOut{Input}{Input}
    \SetKwInOut{Output}{Output}
\Input{$\mathcal{C}$, $\mathcal{D}$, $\mathcal{L}$, $\mathcal{N}$ (with their corresponding cardinalities $C$, $D$, $L$, $N$) /  users' location.}
\Output{Aggregate throughput: ~ $R_{tot}$} 
\DontPrintSemicolon \;
\For{\texttt{\( \mathrm{c}:= 1\) \textbf{to} C}}{
 - allocate random orthogonal RB $\mathrm{n}$ to user $\mathrm{c};$
 $\mathcal{N}_{cellular} = \mathcal{N}_{cellular} - \{\mathrm{n}\};$
}

\For{\texttt{\( \mathrm{i}:= 1\) \textbf{to} D}}{
  \For{\texttt{\( \mathrm{n}:= 1\) \textbf{to} N}}{
    - calculate $\mathbf{d}_{\mathrm{m}}(\mathrm{i}, \mathrm{n});$ \\
    - calculate $\mathbf{r}_{\mathrm{m}}(\mathrm{i}, \mathrm{n});$    
  }
  $\mathbf{d_{\mathrm{m}}^{\mathrm{max}}\mathrm{(i)} ~= ~\mathrm{max}\big(\mathbf{d}_{\mathrm{m}}\mathrm{(i, :)}}\big);$ \\
  $\mathbf{r_{\mathrm{m}}^{\mathrm{max}}\mathrm{(i)} ~= ~\mathrm{max}\big(\mathbf{r}_{\mathrm{m}}\mathrm{(i, :)}}\big);$
}
\DontPrintSemicolon \;
$\mathbf{S} = zeros(D, 2);$ \\
$\mathrm{j} = 1;$ \\
\While{$ \mathrm{j} \leq D$}{
    \textbf{find} $<\mathrm{d}, \mathrm{n}>$ combination that gives the maximum rate among all elements in $\mathbf{d}_{\mathrm{m}}^{\mathrm{max}}$ and $\mathbf{r}_{\mathrm{m}}^{\mathrm{max}}$ matrices$;$ \\
    $\mathbf{S}(\mathrm{d}, :) = [\mathrm{d}, \mathrm{n}];$ \\
    \Repeat{all matrices' rows are updated}{
      - update the rates on the assigned RB $\mathrm{n}$ $\forall \mathrm{u} \in \mathcal{D} - \{\mathrm{d}\}$ for both $\mathbf{d}_{\mathrm{m}}$, $\mathbf{r}_{\mathrm{m}};$ \\
      - update $\mathbf{d}_{\mathrm{m}}^{\mathrm{max}}\mathrm{(u)} ~\&~ \mathbf{r}_{\mathrm{m}}^{\mathrm{max}}\mathrm{(u)};$ \\
    $\mathbf{d}_{\mathrm{m}}(\mathrm{d}, :) = 0; ~ \mathbf{r}_{\mathrm{m}}(\mathrm{d}, :) = 0;$  
    }
    $\mathrm{j} = \mathrm{j} + 1;$
  }
  \For{\texttt{\( \mathrm{d}:= 1\) \textbf{to} D}}{
  - calculate achieved rate for direct or relayed D2D comm. for user $\mathrm{d} ~(R_d);$
  }
  \For{\texttt{\( \mathrm{c}:= 1\) \textbf{to} C}}{
  - calculate achieved rate for cellular link $\mathrm{c} ~(R_c);$
  }
  $R_{tot} = \sum_{c \in \mathcal{C}} R_c + \sum_{d \in \mathcal{D}} R_d;$

  \caption{Sum-rate maximization algorithm}
\end{algorithm}

\section{Heuristic Algorithm}
\label{heuristic}
Herein, an algorithm that prioritizes the D2D users to achieve the maximum rate performance with respect to cellular throughput is devised. A basic assumption is that cellular users are initially allocated with orthogonal resources to satisfy their UL transmissions. Further, we iterate over all D2D links and pre-calculate for each one of them their potential rate performance (according to Shannon capacity formula) on each RB, based on the interference from cellular UEs. Then, we identify the best combination of D2D UE and RB that gives the maximum among all rate as a starting point. Recall that the maximum rate of a UE on a specific RB can result from either direct or relayed communication. Then, we update the rate matrices ($\mathbf{d}_{\mathrm{m}}$ for direct and $\mathbf{r}_{\mathrm{m}}$ for relayed transmission) with the former step's allocation and iterate over all UEs by taking into account the interference deriving from this RB assignment. Last, after all UEs are served, we estimate the rate that each UE achieves through the final allocation pattern and consequently the overall throughput. The algorithmic steps are analytically shown in Algorithm \ref{alg:heuristic_alg}.

\section{Evaluation results}
\label{results}
In this section, a set of numerical investigations is presented to evaluate the performance of the GA-based resource allocation method. The results derive from  Monte Carlo simulations of 100 iterations, implemented in Matlab. Also, one RB is assumed to be assigned for each transmission. The rest of the system parameters are shown in Table \ref{table:simul_param}.

\begin{table}
\centering
\caption{Simulation Parameters }
\begin{tabular}{l r}
\toprule
\textbf{Parameter} & \textbf{Value}\\
\midrule
User distribution & Uniform\\
Macro cell radius  & \(250\) m\\
D2D link length  & \([20, 150]\) m\\
Number of CUEs in cell  & 30 \\
Number of relays/D2D links  & 50 \\
Path-Loss  model & \(128.1 + 37.6\log_{10}d\)  \\ 
UE/relay Tx power (\textit{fixed}) & \(20\) dBm \\
Noise power spectral density & \(-174\) dBm/Hz \\
System bandwidth (\(BW\)) & \(10\) MHz\\
\bottomrule
\end{tabular}
\label{table:simul_param}
\end{table} 

We compare the proposed GA techniques (one-point (OP) and two-points (TP) crossover) with the heuristic RA algorithm that was described in Section \ref{heuristic} and a random RA method. The random method works as follows: after the allocation of orthogonal RBs to cellular UEs takes place, DUEs are also randomly assigned resources from the available RB pool and satisfy their transmission needs by selecting either relay or direct mode, depending on which of the two modes provides better rate performance. 

\begin{figure}
\centering
\includegraphics[width=0.47\textwidth, trim = .5cm .05cm .8cm .5cm,clip = true]{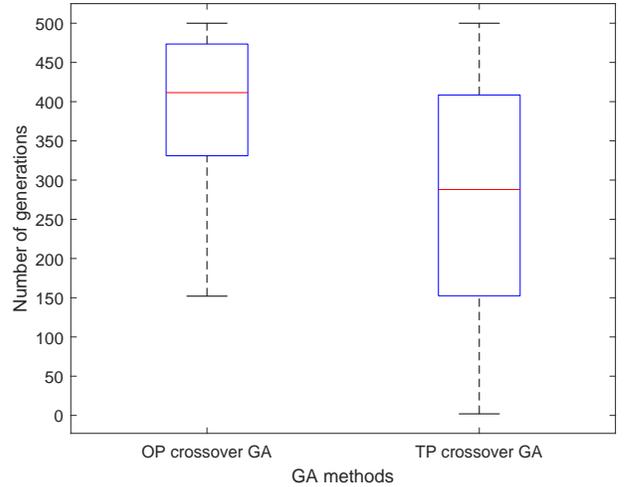}
\caption{Average convergence points for the case of (i) one-point (OP) crossover GA, and (ii) two-points (TP) crossover GA.}
\label{fig:avg_conv}
\end{figure}

\begin{figure}
\centering
\includegraphics[width=0.47\textwidth, trim = .5cm .02cm .8cm .2cm,clip = true]{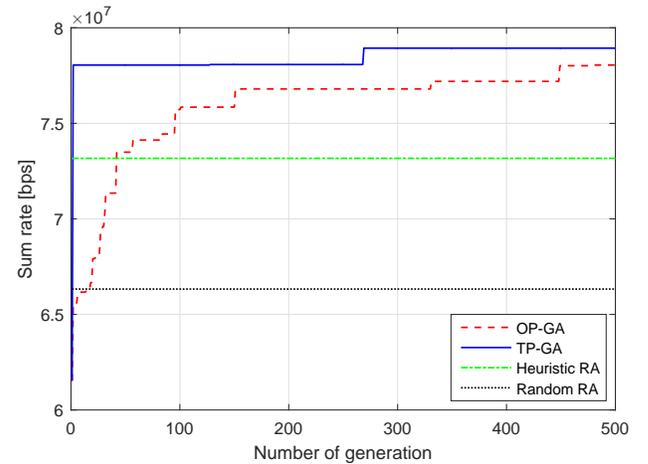}
\caption{An example of the GA's convergence to the maximum rate values.}
\label{fig:sample_conv}
\end{figure}

First, a significant factor that needs to be taken into account is the convergence point of the applied GA methods. This point can be interpreted as the number of generations that results in the optimal achievable aggregate rate. The box plot in Fig. \ref{fig:avg_conv} shows that the TP-GA technique converges almost 1.5 times faster compared to the OP-GA (the medians of convergence points in relation to the number of generations are 290 and 412, respectively). Also, the horizontal edges of each box (25 and 75 percentiles) show a bigger gap in the second case where the TP-GA can achieve a really fast convergence on average. This can be justified by the TP crossover's ability to ensure a more diverse initial population and encoding that can entail faster convergence to the optimal rate. 

Further, Fig. \ref{fig:sample_conv} shows a sample of the sum-rate performance tendency for a designated number of generations. In this case, the TP-GA not only converges faster to its optimal solution (i.e. 210 generations less needed) but also the achievable rate is notably high compared to the heuristic (almost 10\%) and clearly better than the OP-GA method. It has to be noted that, in this case study, the TP-GA method provides a higher capacity performance even from the second generation and beyond, while OP-GA converges in its optimal point in the $468^{\text{th}}$ generation but with rather sub-optimal throughput. Last, for this simulation run, TP-GA outperforms the random method with almost 21\% gain in terms of sum rate performance.

\begin{figure}
\centering
\includegraphics[width=0.47\textwidth, trim = .5cm .04cm .8cm .2cm,clip = true]{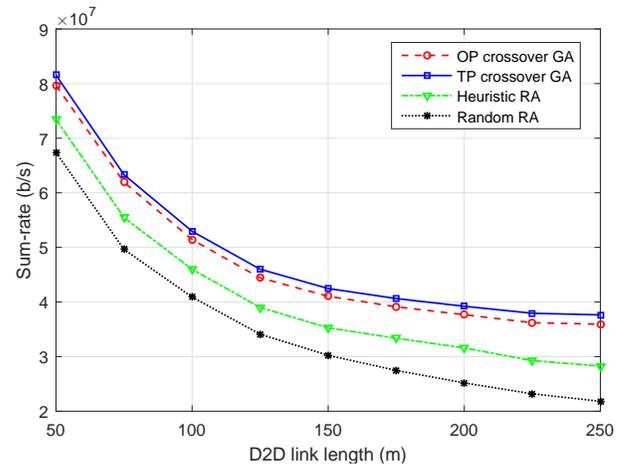}
\caption{Aggregate throughput in relation to varying D2D link lengths.}
\label{fig:rate}
\end{figure}

Fig. \ref{fig:rate} illustrates the sum-rate performance of the proposed methods when the D2D transmitter and receiver are separated by fixed distances for each evaluation point. The TP crossover GA method achieves an average sum-rate gain of 4\%, 24\% and 43\% compared to the OP-GA, heuristic and random allocation techniques, respectively. The plot shows that even though the rate drops proportionally with the increase of the D2D link range, the performance gap of the GA proposed algorithms in comparison to the two RA schemes becomes larger. For the case that the D2D link length is fixed at 250 meters for all DUEs, the TP-GA method provides a rate improvement of 37\% and 72\% compared to the heuristic and the random methods, respectively, signifying a more efficient resource and mode (direct, relayed) selection for D2D communications.       

\begin{figure}
\centering
\includegraphics[width=0.47\textwidth, trim = .5cm .05cm .8cm .2cm,clip = true]{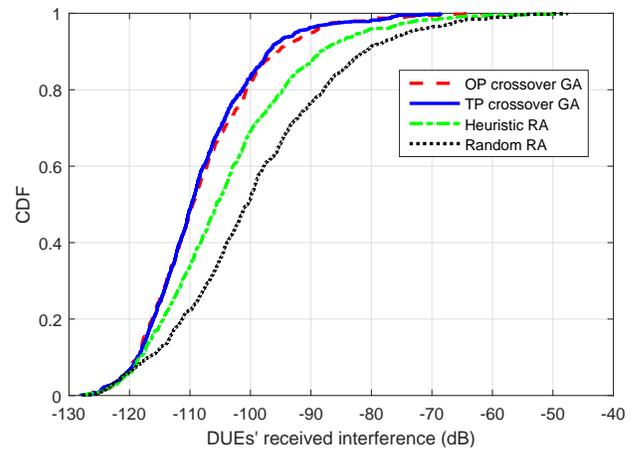}
\caption{CDF of DUEs' received interference.}
\label{fig:rec_interference}
\end{figure}

Finally, we investigate the received interference by D2D UEs for all the considered cases. 
Note that, this interference can result from both a cellular and other D2D/relay transmissions that reuse the same spectrum. 
As shown in Fig. \ref{fig:rec_interference}, the GA methods achieve a lower interference level where at the 50th percentile, the interference level in GA is 4.7 and 10 dB lower than the heuristic and random methods respectively, and at the 90th percentile the GA interference reduction is 9.4 and 15.7 dB compared to the baseline methods.

\section{Conclusions}
\label{conclusion}
In this paper, we presented a resource allocation methodology for relay-aided D2D communications that underlay a cellular network. By exploiting the robustness and versatility of bio-inspired meta-heuristic techniques, we proposed a low-complexity genetic algorithmic framework that aimed at maximizing the network throughput performance with respect to interference. Numerical evaluation results highlight the merits of the investigated GA methods. The proposed one-point and two-points crossover GA techniques provide significant rate improvement amounting to more than 20\% and 40\% compared with heuristic and random RA methods respectively. The proposed GA methods also ensure the least exerted interference towards D2D transmissions with an average gain of more than 4 and 9 dB as compared to the baseline techniques. 
As a future step, power management on top of an efficient GA-based scheme will be studied. In addition, optimization of the relay selection problem will be further investigated. 

\section*{Acknowledgment}
This work is co-funded by Vodafone Group R\&D and CROSSFIRE MITN Marie Curie project (FP7-317126).

\ifCLASSOPTIONcaptionsoff
  \newpage
\fi

\bibliographystyle{IEEEtran}
\bibliography{references}

\end{document}